# On Achievable Rates and Complexity of LDPC Codes for Parallel Channels: Information-Theoretic Bounds and Applications

Igal Sason    Gil Wiechman
Technion, Haifa 32000, Israel
{sason@ee, igillw@tx}.technion.ac.il

*Abstract* — **The paper presents bounds on the achievable rates and the decoding complexity of low-density parity-check (LDPC) codes. It is assumed that the communication of these codes takes place over statistically independent parallel channels where these channels are memoryless, binary-input and output-symmetric (MBIOS). The bounds are applied to punctured LDPC codes. A diagram concludes our discussion by showing interconnections between the theorems in this paper and some previously reported results.**

## I. Introduction

This paper focuses primarily on the information-theoretic limitations of low-density parity-check (LDPC) codes whose transmission takes place over a set of parallel channels. By parallel channels we mean a communication system or model where the distortion introduced by each channel is independent of the signal and the distortion in all of the other channels. The first main result presented in this paper is the derivation of an upper bound on the achievable rates of ensembles of LDPC codes under optimal maximum-likelihood (ML) decoding when we assume that the codes are communicated over parallel memoryless binary-input output-symmetric (MBIOS) channels. This result forms a non-trivial generalization of the "un-quantized" bound derived by the authors for a single MBIOS channel [11], where the latter bound improves the tightness of the bound in [1] which was based on a two-level quantization approach. The second main result introduced in this paper is a derivation of a lower bound on the decoding complexity of ensembles of LDPC codes for parallel MBIOS channels; this bound assumes an iterative message-passing decoder, and the decoding complexity is normalized per iteration. In this respect, it continues the study in [2] which was focused on upper bounds on the parity-check density of ensembles of LDPC codes for parallel MBIOS channels. The two main issues addressed in this paper for parallel channels are applied to punctured LDPC codes. The bounds on the achievable rates and decoding complexity of LDPC codes over statistically independent parallel channels are applied to the case of transmission of ensembles of punctured LDPC codes over a single MBIOS channel.

The interested reader is referred to the full paper version [10] which provides proofs to all of the theorems presented here, and further discusses the relations between these theorems and some other previously reported results from [1, 5, 6, 8, 9, 11].

## II. An Upper bound on the Achievable Rates of LDPC codes over Parallel Channels

In this section, we present an upper bound on the design rate of a sequence of ensembles of LDPC codes whose transmission takes place over a set of statistically independent parallel MBIOS channels, and achieves vanishing bit error probability under ML decoding. This bound is used in the next section for the derivation of an upper bound on the design rate of an arbitrary sequence of ensembles of punctured LDPC codes.

Let us assume that a binary LDPC code $\mathcal{C}$ of length $n$ is transmitted over a set of $J$ statistically independent parallel MBIOS channels. Denote the number of code bits of $\mathcal{C}$ which are transmitted over the $j^{\text{th}}$ channel by $n^{[j]}$, and the fraction of bits transmitted over the $j^{\text{th}}$ channel by

$$p_j \triangleq \frac{n^{[j]}}{n}, \quad j \in \{1, \ldots, J\}. \qquad (1)$$

Let $\mathcal{G}$ be a bipartite graph which represents the code $\mathcal{C}$, and $E$ be the set of edges in $\mathcal{G}$. Let $E^{[j]}$ designate the set of edges connected to variable nodes which correspond to code bits transmitted over the $j^{\text{th}}$ channel, and

$$q_j \triangleq \frac{|E^{[j]}|}{|E|}, \quad j \in \{1, \ldots, J\} \qquad (2)$$

denote the fraction of edges connected to these variable nodes. Referring to the edges from the subset $E^{[j]}$, let $\lambda_i^{[j]}$ designate the fraction of these edges which are connected to variable nodes of degree $i$, and define the following $J$ degree distributions from the edge perspective:

$$\lambda^{[j]}(x) \triangleq \sum_{i=2}^{\infty} \lambda_i^{[j]} x^{i-1}, \qquad j \in \{1, \ldots, J\}$$

which correspond to each of the $J$ parallel channels. For the simplicity of the notation, let us define a vector of degree distributions for the variable nodes from the edge perspective to be $\underline{\lambda}(x) = \left(\lambda^{[1]}(x), \ldots, \lambda^{[J]}(x)\right)$. Following the notation in [6], the ensemble $(n, \underline{\lambda}, \rho)$ is defined to be the set of LDPC codes of length $n$, which according to their representation by bipartite graphs and the assignment of their code bits to the parallel channels, imply left and right degree distributions of $\underline{\lambda}$ and $\rho$, respectively.

In the following, we introduce a sequence of ensembles of LDPC codes, say $\{(n_r, \underline{\lambda}_r, \rho)\}_{r=1}^{\infty}$ where the fraction of code bits assigned to each of the $J$ parallel channels is uniform over all the codes of each ensemble, and $\rho$ is fixed for all the ensembles of this sequence (i.e., it is independent of $r$). Under the assumption that $\lambda$, which corresponds to the overall left degree distribution of the edges, is independent of $r$, one can consider here the common design rate of the sequence of ensembles $\{(n_r, \underline{\lambda}_r, \rho)\}_{r=1}^{\infty}$ which does not depend on $r$.

This setting is general enough for applying the following theorem to various applications which form particular cases of communication over parallel channels. In this setting, the fraction of code bits assigned to the $j^{\text{th}}$ channel, $p_{j,r}$, depends on $j \in \{1, \ldots, J\}$ and $r \in \mathbb{N}$, but not on the particular code chosen from each ensemble. It follows from [10, Lemma 4.2] that the same property also holds for $q_{j,r}$ which designates the fraction of edges which are connected to variable nodes whose code bits are assigned to the $j^{\text{th}}$ channel. In the following, we assume that the limits

$$p_j \triangleq \lim_{r \to \infty} p_{j,r}, \quad q_j \triangleq \lim_{r \to \infty} q_{j,r} \qquad (3)$$

exist and are positive for all $j \in \{1, \ldots, J\}$.

**Theorem II.1** Let a sequence of LDPC ensembles $\{(n_r, \underline{\lambda}_r, \rho)\}_{r=1}^{\infty}$ be transmitted over a set of $J$ statistically independent parallel MBIOS channels, and assume that the block length $(n_r)$ goes to infinity as we let $r$ tend to infinity. Let $C_j$ denote the capacity of the $j^{\text{th}}$ channel, and $a(\cdot; j)$ designate the *pdf* of the LLR at the output of the $j^{\text{th}}$ channel given its input is 1 (where the input alphabet is $\{+1, -1\}$). If in the limit where $r$ tends to infinity, the bit error probability of this sequence under ML decoding vanishes, then the common design rate $R_d$ of these ensembles satisfies

$$R_d \leq 1 - \frac{1 - \sum_{j=1}^{J} p_j C_j}{1 - \frac{1}{2\ln 2} \sum_{p=1}^{\infty} \left\{ \frac{1}{p(2p-1)} \Gamma\left(\sum_{j=1}^{J} q_j \, g_{j,p}\right) \right\}} \qquad (4)$$

where $\Gamma$ denotes the right degree distribution from the node perspective, and for all $j \in \{1, \ldots, J\}$ and $p \in \mathbb{N}$, $g_{j,p}$ is defined as follows:

$$g_{j,p} \triangleq \int_0^{\infty} a(l; j) \, (1 + e^{-l}) \tanh^{2p}\left(\frac{l}{2}\right) dl. \qquad (5)$$

**Example II.1** For the particular case where the $J$ parallel MBIOS channels are binary erasure channels where the erasure probability of the $j^{\text{th}}$ channel is $\varepsilon_j$, we get from (5)

$$g_{j,p} = 1 - \varepsilon_j, \quad \forall \, j \in \{1, \ldots, J\}, \ p \in \mathbb{N}. \qquad (6)$$

Since $g_{j,p}$ is independent of $p$ for a BEC, and based on the equality $\sum_{p=1}^{\infty} \frac{1}{2p(2p-1)} = \ln 2$, we obtain from Theorem II.1 that the common design rate of the sequence of LDPC ensembles is upper bounded by

$$R_d \leq 1 - \frac{\sum_{j=1}^{J} p_j \varepsilon_j}{1 - \Gamma\left(1 - \sum_{j=1}^{J} q_j \, \varepsilon_j\right)}. \qquad (7)$$

This particular result coincides with [6, Theorem 2].

## III. On the Achievable Rates of Punctured LDPC Codes

In this section we present upper bounds on the achievable rates of punctured LDPC codes whose transmission takes place over an MBIOS channel, and the codes are ML decoded. The transmission of punctured codes can be interpreted as a special case of transmitting the original (un-punctured) codes over a set of parallel channels where these component channels are formed by a mixture of the communication channel and BECs whose erasure probabilities are the puncturing rates of different subsets of code bits. Hence, the analysis in this section relies on the bound presented in Section II.

**Randomly Punctured LDPC Codes** In this section, we consider the achievable rates of randomly punctured LDPC (RP-LDPC) codes. We assume that the transmission of these codes takes place over an MBIOS channel, and refer to their achievable rates under optimal ML decoding. The upper bound on the achievable rates of ensembles of RP-LDPC codes relies on Section II where we presented an upper bound on the achievable rates of LDPC codes for parallel channels.

In the following, we assume that the communication takes place over an MBIOS channel with capacity $C$, and we define

$$g_p \triangleq \int_0^{\infty} a(l) \, (1 + e^{-l}) \tanh^{2p}\left(\frac{l}{2}\right) dl, \qquad p \in \mathbb{N} \qquad (8)$$

where $a$ designates the *pdf* of the LLR of the channel given that its input is zero.

**Theorem III.1** Let $\{(n_r, \lambda, \rho)\}_{r=1}^{\infty}$ be a sequence of ensembles of LDPC codes whose block length $(n_r)$ tends to infinity as $r \to \infty$. Assume that a sequence of ensembles of RP-LDPC codes is constructed in the following way: for each code from an ensemble of the original sequence, a subset of $\alpha n_r$ code bits is a-priori selected, and these bits are randomly punctured at a fixed rate $(P_{\text{pct}})$. Assume that the punctured codes are transmitted over an MBIOS channel with capacity $C$, and that in the limit where $r$ approaches infinity, the sequence of ensembles of RP-LDPC codes achieves vanishing bit error probability under some decoding algorithm. Then in probability 1 w.r.t. the random puncturing patterns, the asymptotic design rate $(R_d)$ of the new sequence satisfies

$$R_d \leq \frac{1}{1 - \alpha P_{\text{pct}}} \left(1 - \frac{1 - (1 - \alpha P_{\text{pct}})C}{1 - \frac{1}{2\ln 2} \sum_{p=1}^{\infty} \left\{\frac{1}{p(2p-1)} \Gamma\left((1 - P_{\text{pct}} + \xi)g_p\right)\right\}}\right) \qquad (9)$$

where $\Gamma$ denotes the right degree distribution (from the node perspective) of the original sequence, $g_p$ is introduced in (8), and $\xi$ is the following positive number:

$$\xi \triangleq 2(1-\alpha)P_{\text{pct}} \int_0^1 \lambda(x) \, dx. \qquad (10)$$

**Intentionally Punctured LDPC Codes** In [3], Ha and McLaughlin show that good codes can be constructed by puncturing good ensembles of LDPC codes using a technique called "intentional puncturing". In this approach, the code bits are partitioned into disjoint sets so that each set contains all the code bits whose corresponding variable nodes have the same degree. The code bits in each one of these sets are randomly punctured at a fixed puncturing rate.

We briefly present the notation used in [3] for the characterization of ensembles of intentionally punctured LDPC (IP-LDPC) codes. Consider an ensemble of LDPC codes with left and right edge degree distributions $\lambda$ and $\rho$, respectively. For each degree $j$ such that $\lambda_j > 0$, a puncturing rate $\pi_j \in [0, 1]$ is determined for randomly puncturing the set of code bits which correspond to variable nodes of degree $j$. The polynomial associated with this puncturing pattern is

$$\pi^0(x) \triangleq \sum_{j=1}^{\infty} \pi_j x^{j-1}. \qquad (11)$$

An ensemble of IP-LDPC codes can be therefore represented by the quadruplet $(n, \lambda, \rho, \pi^0)$ where $n$ designates the block length of these codes, $\lambda$ and $\rho$ are the left and right degree distributions from the edge perspective, respectively, and $\pi^0$ is the polynomial which corresponds to the puncturing pattern, as given in (11). The average fraction of punctured bits is given by $p^{(0)} = \sum_{j=1}^{\infty} \Lambda_j \pi_j$ where $\Lambda$ is the left node degree distribution of the original LDPC ensemble. The following statement, which relies on Theorem II.1, provides an upper bound on the common design rate of a sequence of ensembles of IP-LDPC codes. This bound refers to ML decoding (and hence, to any sub-optimal decoding algorithm).

**Theorem III.2** Let $\{(n_r, \lambda, \rho, \pi^0)\}_{r=1}^{\infty}$ be a sequence of ensembles of IP-LDPC codes transmitted over an MBIOS channel, and assume that $n_r$ tends to infinity as $r \to \infty$. Let $C$ be the channel capacity, and $a$ be the *pdf* of the LLR at the output of the channel given its input is 1. If the asymptotic bit error probability of this sequence vanishes under ML decoding (or any sub-optimal decoding algorithm) as $r \to \infty$, then in probability 1 w.r.t. the puncturing patterns, the common design rate $R_\mathrm{d}$ of these ensembles satisfies

$$R_\mathrm{d} \leq \frac{1}{1-p^{(0)}} \left[ 1 - \frac{1-(1-p^{(0)})C}{1 - \frac{1}{2\ln 2} \sum_{p=1}^{\infty} \left\{ \frac{1}{p(2p-1)} \, \Gamma\!\left( \left(1 - \sum_{j=1}^{\infty} \lambda_j \pi_j \right) g_p \right) \right\}} \right] \quad (12)$$

where $\Gamma$ denotes the right degree distribution from the node perspective,

$$p^{(0)} \triangleq \sum_{j=1}^{\infty} \Lambda_j \pi_j \quad (13)$$

designates the average puncturing rate of the code bits, and $g_p$ is the functional of the MBIOS channel introduced in (8).

**Numerical Results for Intentionally Punctured LDPC Codes** In this section, we compare between thresholds under message-passing iterative (MPI) decoding and bounds on thresholds under ML decoding for ensembles of IP-LDPC codes. It is assumed that the transmission of the punctured LDPC codes takes place over a binary-input AWGN channel. The pairs of degree distributions and the corresponding puncturing patterns were originally presented in [3]. We use these ensembles in order to study their inherent gap to capacity, and also study how close to optimal is iterative decoding for these ensembles (in the asymptotic case where the block length goes to infinity). Table 1 refers to an ensemble of rate$-\frac{1}{2}$ LDPC codes which by puncturing, the rate varies between 0.50 and 0.91.

Table 1 provides lower bounds on the inherent gap to capacity under optimal ML decoding (based on Theorem III.2); these values are compared to the corresponding gaps to capacity under iterative message-passing decoding (whose calculation is based on the density evolution analysis). On one hand, Table 1 provides a quantitative assessment of the loss in the asymptotic performance which is attributed to the sub-optimality of iterative decoding (as compared to optimal ML decoding), and on the other hand, they provide an assessment of the inherent loss in performance which is attributed to the structure of the ensembles, even if optimal ML decoding could be applied to decode these codes. The loss in performance in both cases is measured in terms of $\frac{E_\mathrm{b}}{N_0}$ in decibels. It is demonstrated in Table 1 that the asymptotic loss in performance due to the code structure is still non-negligible as compared to the corresponding loss due to the sub-optimality of iterative decoding. For all the ensembles of IP-LDPC codes considered in Table 1 (which were originally introduced in [3, Table 2]), the gap to capacity under the sum-product iterative decoding algorithm does not exceed 0.6 dB; however, under ML decoding, the gap to capacity is always greater than $\frac{1}{3}$ of the corresponding gap to capacity under this iterative decoding algorithm; therefore, the results in Table 1 regarding the thresholds under ML decoding further emphasize the efficiency of the sum-product decoding algorithm for these ensembles, especially in light of its moderate complexity. Table 1 also shows that the performance of the punctured LDPC codes is degraded at high rates, where one needs to pay a considerable penalty for using punctured codes. This phenomenon was explained in [7, Theorem 1] by the threshold effect for ensembles of IP-LDPC codes.

## IV. On the Decoding Complexity of LDPC Codes for Parallel Channels

We present in this section a lower bound on the decoding complexity of LDPC codes for parallel MBIOS channels and lower bounds on the decoding complexity of punctured LDPC codes, and later (see Section V) relate the new bounds to previously reported results.

**A Lower Bound on the Decoding Complexity for Parallel MBIOS Channels** Consider a binary linear block code which is represented by a bipartite graph, and assume that the graph serves for the decoding with an iterative algorithm. Following [2] and [5], the decoding complexity under MPI decoding is defined as the number of edges in the graph normalized per information bit. This quantity measures the number of messages which are delivered through the edges of the graph (from left to right and vice versa) during a single iteration. Equivalently, since there is a one-to-one correspondence between a bipartite graph and the parity-check matrix $H$ which represents the code, the decoding complexity is also equal to the number of non-zero elements in $H$ normalized per information bit (i.e., the density of the parity-check matrix [9, Definition 2.2]). Hence, the decoding complexity (as well as the performance) of iteratively decoded binary linear block codes depends on the specific representation of the code by a parity-check matrix. Since the average right degree ($a_\mathrm{R}$) of a bipartite graph is equal to the number of edges normalized per parity-check equation, then the average right degree and the decoding complexity are related quantities. Consider an ensemble of LDPC codes whose design rate is $R_\mathrm{d}$. It is natural to relate the decoding complexity of the ensemble, say $\chi_D$, to its average right degree and design rate, as follows:

$$\chi_\mathrm{D} = \frac{1-R_\mathrm{d}}{R_\mathrm{d}} \, a_\mathrm{R} \;.$$

We note that $a_\mathrm{R}$ is fixed for all the codes from an ensemble of LDPC codes with a given pair of degree distributions.

Consider a sequence of ensembles of LDPC codes, $\{(n_r, \underline{\lambda}_r, \rho)\}_{r=1}^{\infty}$, whose transmission takes place over a set of $J$ statistically independent parallel MBIOS channels. Let $C_j$ and $p_j$ be the capacity and the fraction of code bits which are assigned to the $j^{\mathrm{th}}$ channel, respectively (where $j \in \{1, \ldots, J\}$). We now present a lower bound on the decoding complexity per iteration under MPI decoding for this sequence.

| $\pi^0(x)$ (puncturing pattern) | Design rate | Capacity limit | Lower bound (ML decoding) | Iterative (IT) Decoding | Fractional gap to capacity (ML vs. IT) |
|---|---|---|---|---|---|
| 0 | 0.500 | 0.187 dB | 0.270 dB | 0.393 dB | $\geq 40.3\%$ |
| $0.07886x + 0.01405x^2 + 0.06081x^3 + 0.07206x^9$ | 0.528 | 0.318 dB | 0.397 dB | 0.526 dB | $\geq 37.9\%$ |
| $0.20276x + 0.09305x^2 + 0.03356x^3 + 0.16504x^9$ | 0.592 | 0.635 dB | 0.716 dB | 0.857 dB | $\geq 36.4\%$ |
| $0.25381x + 0.15000x^2 + 0.34406x^3 + 0.019149x^9$ | 0.629 | 0.836 dB | 0.923 dB | 1.068 dB | $\geq 37.3\%$ |
| $0.31767x + 0.18079x^2 + 0.05265x^3 + 0.24692x^9$ | 0.671 | 1.083 dB | 1.171 dB | 1.330 dB | $\geq 35.6\%$ |
| $0.36624x + 0.24119x^2 + 0.49649x^3 + 0.27318x^9$ | 0.719 | 1.398 dB | 1.496 dB | 1.664 dB | $\geq 36.9\%$ |
| $0.41838x + 0.29462x^2 + 0.05265x^3 + 0.30975x^9$ | 0.774 | 1.814 dB | 1.927 dB | 2.115 dB | $\geq 37.2\%$ |
| $0.47074x + 0.34447x^2 + 0.02227x^3 + 0.34997x^9$ | 0.838 | 2.409 dB | 2.547 dB | 2.781 dB | $\geq 37.1\%$ |
| $0.52325x + 0.39074x^2 + 0.01324x^3 + 0.39436x^9$ | 0.912 | 3.399 dB | 3.607 dB | 3.992 dB | $\geq 35.1\%$ |

Table 1: Comparison of thresholds for ensembles of IP-LDPC codes where the original ensemble before puncturing has the degree distributions $\lambda(x) = 0.25105x + 0.30938x^2 + 0.00104x^3 + 0.43853x^9$ and $\rho(x) = 0.63676x^6 + 0.36324x^7$ (so its design rate is equal to $\frac{1}{2}$). The transmission of these codes takes place over a binary-input AWGN channel. The table compares values of $\frac{E_b}{N_0}$ referring to the capacity limit, the bound given in Theorem III.2 (which provides a lower bound on $\frac{E_b}{N_0}$ under ML decoding), and thresholds under iterative message-passing decoding. The fractional gap to capacity (see the rightmost column) measures the ratio of the gap to capacity under optimal ML decoding and the achievable gap to capacity under (sub-optimal) iterative message-passing decoding. The pair of degree distributions for the ensemble of LDPC codes, and the polynomials which correspond to its puncturing patterns are taken from [3, Table 2].

**Theorem IV.1** Let a sequence of ensembles of LDPC codes, $\{(n_r, \underline{\lambda}_r, \rho)\}_{r=1}^{\infty}$, be transmitted over a set of $J$ statistically independent parallel MBIOS channels. Assume that the capacities $C_j$ of these channels are all positive, and denote the average capacity by $\overline{C} \triangleq \sum_{j=1}^{J} p_j C_j$. If this sequence achieves a fraction $1 - \varepsilon$ of $\overline{C}$ with vanishing bit error probability, then the asymptotic decoding complexity under MPI decoding satisfies

$$\chi_D(\varepsilon) \geq K_1 + K_2 \ln\left(\frac{1}{\varepsilon}\right). \quad (14)$$

The coefficients $K_{1,2}$ in this lower bound are as follows:

$$K_1 = -\frac{(1 - \overline{C}) \ln\left(\frac{1}{2 \ln 2} \frac{1-\overline{C}}{\overline{C}}\right)}{\overline{C} \ln\left(\sum_{j=1}^{J} q_j g_{j,1}\right)}, \quad K_2 = -\frac{1 - \overline{C}}{\overline{C} \ln\left(\sum_{j=1}^{J} q_j g_{j,1}\right)} \quad (15)$$

where $g_{j,1}$ is introduced in (5), and $q_j$ is introduced in (3) and is assumed to be positive for all $j \in \{1, \ldots, J\}$. For parallel BECs, the term $\frac{1}{2 \ln 2}$ can be removed from the numerator in the expression of $K_1$.

**Lower Bounds on the Decoding Complexity for Punctured LDPC Codes** Consider an ensemble of LDPC codes of length $n$ and design rate $R_d'$, and let the code bits be partitioned into $J$ disjoint sets where the $j^{\text{th}}$ set contains a fraction $p_j$ of these bits ($j \in \{1, \ldots, J\}$). Assume that the bits in the $j^{\text{th}}$ set are randomly punctured at rate $\pi_j$, and let the punctured codes be transmitted over an MBIOS channel whose capacity is $C$. As shown in the previous section, this is equivalent to transmitting the original (un-punctured) codes over a set of $J$ parallel channels, where the $j^{\text{th}}$ set of code bits is transmitted over a channel whose capacity is $C_j = (1-\pi_j)C$. The average capacity of this set of $J$ parallel channels is given by $\overline{C} = \sum_{j=1}^{J} p_j (1 - \pi_j) C = (1 - \gamma) C$ where $\gamma \triangleq \sum_{j=1}^{J} p_j \pi_j$ is the overall puncturing rate. Denote the design rate of the punctured codes by $R_d \triangleq \frac{R_d'}{1-\gamma}$, then it follows that the multiplicative gap to capacity of the punctured codes is given by $\varepsilon = 1 - \frac{R_d}{C} = 1 - \frac{R_d'}{\overline{C}}$. In the following, we present a lower bound on the decoding complexity of RP-LDPC codes.

**Theorem IV.2** Let $\{(n_r, \lambda, \rho)\}_{r=1}^{\infty}$ be a sequence of ensembles of LDPC codes whose block length ($n_r$) tends to infinity as $r \to \infty$. Assume that a sequence of ensembles of RP-LDPC codes is constructed in the following way: for each code from an ensemble of the original sequence, a subset of $\alpha n_r$ code bits is a-priori selected, and these bits are randomly punctured at a fixed rate ($P_{\text{pct}}$). Assume that the punctured codes are transmitted over an MBIOS channel with capacity $C$, and that as $r$ tends to infinity, the sequence of ensembles of punctured codes achieves a fraction $1 - \varepsilon$ of the capacity with vanishing bit error probability. Then in probability 1 w.r.t. the random puncturing patterns, the decoding complexity of this sequence under MPI decoding satisfies

$$\chi_D(\varepsilon) \geq K_1 + K_2 \ln\left(\frac{1}{\varepsilon}\right). \quad (16)$$

The coefficients $K_{1,2}$ in this lower bound are are given explicitly in [10, Eq. (72)].

The upper bound on the decoding complexity for sequences of ensembles of IP-LDPC codes is also given in terms of the gap between the rate of the punctured rate and the channel capacity.

**Theorem IV.3** Let $\{(n_r, \lambda, \rho, \pi^0)\}_{r=1}^{\infty}$ be a sequence of ensembles of IP-LDPC codes transmitted over an MBIOS channel whose capacity is $C$. If this sequence achieves a fraction $1-\varepsilon$ of the capacity with vanishing bit error probability, then in probability 1 w.r.t. the random puncturing patterns, the decoding complexity of this sequence under MPI decoding satisfies

$$\chi_{\mathrm{D}}(\varepsilon) \geq K_1 + K_2 \ln\left(\frac{1}{\varepsilon}\right). \qquad (17)$$

The coefficients $K_{1,2}$ in this lower bound are given explicitly in [10, Eq. (75)].

## V. Summary and Outlook

The main result in this paper, Theorem II.1, provides an upper bound on the asymptotic rate of a sequence of ensembles of LDPC codes which achieves vanishing bit error probability. We assume that the communication takes place over a set of parallel memoryless binary-input output-symmetric (MBIOS) channels. The derivation of Theorem II.1 relies on upper and lower bounds on the conditional entropy of the transmitted codeword given the received sequence at the output of the parallel channels (see [10, Section 3]), and it is valid under optimal ML decoding (or any sub-optimal decoding algorithm). This theorem enables the derivation of a lower bound on the decoding complexity (per iteration) of ensembles of LDPC codes under message-passing iterative decoding when the transmission of the codes takes place over parallel MBIOS channels. The latter bound is given in terms of the gap between the rate of these codes for which reliable communication is achievable and the channel capacity. Similarly to a lower bound on the decoding complexity of ensembles of LDPC codes for a single MBIOS channel [9], the lower bound on the decoding complexity which is presented for parallel channels also grows like the log of the inverse of the gap to capacity.

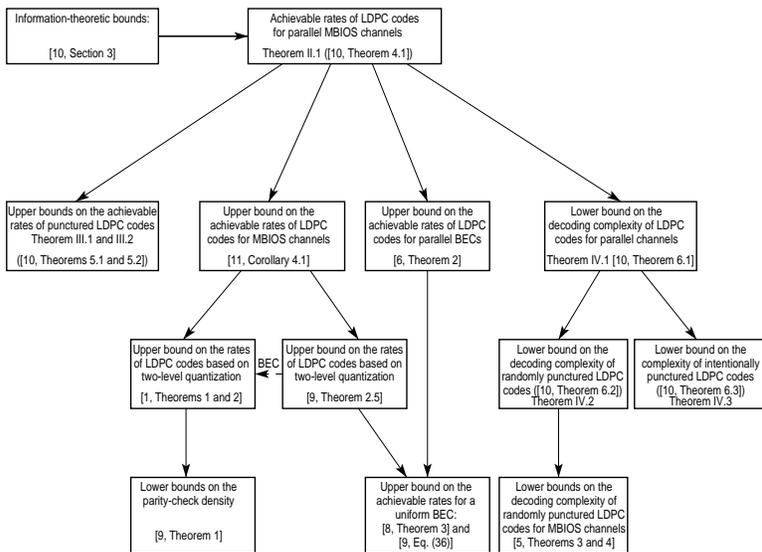

Figure 1: An interconnections diagram among the bounds in this paper and previously reported bounds.

Theorem II.1 can be used for various applications which form particular cases of communication over parallel channels, e.g., intentionally punctured LDPC codes [3], non-uniformly error protected LDPC codes [6], and LDPC-coded modulation schemes. In [10, Section 5], we use Theorem II.1 for the derivation of upper bounds on the achievable rates under ML decoding of (randomly and intentionally) punctured LDPC codes whose transmission takes place over an MBIOS channel. It is exemplified numerically that for various good ensembles of IP-LDPC codes, the asymptotic loss in performance due to the code structure is still non-negligible as compared to the corresponding loss due to the sub-optimality of iterative decoding (as compared to optimal ML decoding). Looser versions of the bounds presented here for punctured LDPC codes suggest a simplified re-derivation of previously reported bounds on the decoding complexity of randomly punctured LDPC codes (see [10]). Interconnections between the theorems introduced in this paper and some other previously reported results are shown in Fig. 1. The proofs of all the theorems in this paper and further discussion on the theorems are provided in [10].

**Acknowledgment** I. Sason wishes to acknowledge Henry Pfister and Ruediger Urbanke for stimulating discussions on puncturing theorems during the preparation of the work in [5]. The work of I. Sason was supported by the Taub and Shalom Foundations.

## References

[1] D. Burshtein, M. Krivelevich, S. Litsyn and G. Miller, "Upper bounds on the rate of LDPC codes," *IEEE Trans. on Information Theory*, vol. 48, no. 9, pp. 2437–2449, September 2002.

[2] Y. Liu, J. Hou and V. K. N. Lau, "Complexity Bounds of LDPC codes for parallel channels," *Proceedings Forty-Second Annual Allerton Conference on Communication, Control and Computing*, pp. 1705–1713, IL, USA, September 2004.

[3] J. Ha and S. W. McLaughlin, "Rate-compatible puncturing of low-density parity-check codes," *IEEE Trans. on Information Theory*, vol. 50, no. 11, pp. 2824–2836, November 2004.

[4] V. Nagarajan, Y. Liu and J. Hou, "Joint design of LDPC codes for parallel channels and its applications," *Proceedings of the Forty-First Annual Allerton Conference on Communication, Control and Computing*, pp. 1704–1705, Urbana-Champaign, IL, USA, October 2003.

[5] H. D. Pfister, I. Sason and R. Urbanke, "Capacity-achieving ensembles for the binary erasure channel with bounded complexity," *IEEE Trans. on Information Theory*, vol. 51, no. 7, pp. 2352–2379, July 2005.

[6] H. Pishro-Nik, N. Rahnavard and F. Fekri, "Nonuniform error correction using low-density parity-check codes," *IEEE Trans. on Information Theory*, vol. 51, no. 7, pp. 2702–2714, July 2005.

[7] H. Pishro-Nik and F. Fekri, "Results on punctured LDPC codes," *Proceedings 2004 IEEE Information Theory Workshop (ITW 2004)*, pp. 215–219, San Antonio, Texas, USA, October 24–29, 2004.

[8] H. Pishro-Nik and F. Fekri, "On decoding of low-density parity-check codes over the binary erasure channel," *IEEE Trans. on Information Theory*, vol. 50, pp. 439–454, March 2004.

[9] I. Sason and R. Urbanke, "Parity-check density versus performance of binary linear block codes over memoryless symmetric channels," *IEEE Trans. on Information Theory*, vol. 49, no. 7, pp. 1611–1635, July 2003.

[10] I. Sason and G. Wiechman, "On achievable rates and complexity of LDPC codes for parallel channels with application to puncturing," submitted to IEEE Trans. on Information Theory, August 2005. See http://arxiv.org/abs/cs.IT/0508072.

[11] G. Wiechman and I. Sason, "Improved bounds on the parity-check density and achievable rates of binary linear block codes with applications to LDPC codes," submitted to *IEEE Trans. on Information Theory*, May 2005. [Online]. Available: `http://arxiv.org/abs/cs.IT/0505057`.